\documentclass{article}
\PassOptionsToPackage{numbers,sort&compress}{natbib}
\usepackage[preprint]{neurips_2025}
\usepackage[T1]{fontenc}
\usepackage{booktabs}
\usepackage{array}
\usepackage{graphicx}
\usepackage{xcolor}
\usepackage{amsmath}
\usepackage{float}
\usepackage{pgfplots}
\pgfplotsset{compat=1.17}
\usepackage{fvextra}
\usepackage[hidelinks]{hyperref}

\newcommand{\bandHigh}{5\%}

\newcommand{\bandZero}{85\%}

\newcommand{\baselineClaudeRecall}{81\%}

\newcommand{\baselineFFR}{6\%}

\newcommand{\baselineGptRecall}{77\%}

\newcommand{\baselineLost}{7}
\newcommand{\baselineRecall}{72\%}

\newcommand{\baselineRecallProFrac}{16/30}

\newcommand{\ctrlFailing}{812}

\newcommand{\ctrlNewOnly}{51\%}

\newcommand{\exJson}{20 of 20}
\newcommand{\exMaxlength}{22 of 23}
\newcommand{\exPylint}{30 of 30}
\newcommand{\flagCands}{2,186}
\newcommand{\flagDeclined}{15}
\newcommand{\flagFailing}{2,434}

\newcommand{\flagFirst}{92}
\newcommand{\flagFlipAll}{61\%}

\newcommand{\flagFlipAllK}{1,331}

\newcommand{\flagFlipLarge}{66\%}
\newcommand{\flagFlipMid}{74\%}

\newcommand{\flagFlipSmall}{77\%}
\newcommand{\flagGainMax}{60}
\newcommand{\flagGainMaxPts}{12}
\newcommand{\flagGainMedian}{45}
\newcommand{\flagGainMin}{33}
\newcommand{\flagGroup}{117}

\newcommand{\flagLinesMedian}{11}
\newcommand{\flagNewOnly}{90\%}

\newcommand{\flagNoFlipK}{22}
\newcommand{\flagRankChanged}{28}
\newcommand{\flagRankMove}{10}
\newcommand{\flagTasks}{108}

\newcommand{\flagTasksWithFlipK}{86}
\newcommand{\flagValid}{93}
\newcommand{\flipDiffCI}{14--50}
\newcommand{\forcedCands}{417}
\newcommand{\forcedDeclined}{1}

\newcommand{\forcedFlipAll}{29\%}

\newcommand{\forcedFlipAllK}{120}

\newcommand{\forcedGainMax}{6}

\newcommand{\forcedGainMin}{2}
\newcommand{\forcedGroup}{296}
\newcommand{\forcedInvalid}{2}
\newcommand{\forcedLinesMedian}{5}
\newcommand{\forcedNoFlipK}{60}

\newcommand{\forcedRest}{8\%}
\newcommand{\forcedRestTasks}{74}
\newcommand{\forcedTasks}{79}

\newcommand{\forcedTopFiveFlips}{96}
\newcommand{\forcedValid}{76}

\newcommand{\freeTasks}{79}

\newcommand{\fullClaudeRecall}{83\%}

\newcommand{\fullFFR}{3\%}

\newcommand{\fullGptRecall}{88\%}

\newcommand{\fullKappa}{0.78}
\newcommand{\fullRecall}{81\%}

\newcommand{\fullRecallProFrac}{22/30}

\newcommand{\fullUncertain}{10}
\newcommand{\goldChanged}{5}

\newcommand{\goldFFR}{5\%}

\newcommand{\goldKappa}{0.87}

\newcommand{\goldRecall}{87\%}

\newcommand{\goldUncertain}{5}
\newcommand{\nAblation}{138}

\newcommand{\nProNarrow}{75}

\newcommand{\nSubmissions}{155}
\newcommand{\nTest}{206}
\newcommand{\nValidation}{344}

\newcommand{\nofindDiffCI}{40--58}

\newcommand{\nofindFlipAll}{49\%}

\newcommand{\nofindTasks}{108}

\newcommand{\nofindValid}{82}

\newcommand{\proFlag}{35\%}

\newcommand{\proFlagFrac}{258/731}
\newcommand{\proFlagK}{258}
\newcommand{\proFlagRest}{31\%}
\newcommand{\proFlagRestFrac}{201/656}

\newcommand{\proScoreFour}{115}

\newcommand{\proTasks}{731}

\newcommand{\relaxCalls}{110}
\newcommand{\relaxTokensIn}{5,081}
\newcommand{\relaxTokensOut}{2,052}
\newcommand{\rerunChanged}{6}

\newcommand{\rerunGained}{2}

\newcommand{\rerunLost}{3}

\newcommand{\testAgree}{91\%}

\newcommand{\testFFR}{4\%}

\newcommand{\testFalseFlags}{4}
\newcommand{\testGptFFR}{18\%}

\newcommand{\testGptRecall}{94\%}

\newcommand{\testKappa}{0.82}
\newcommand{\testRecall}{89\%}

\newcommand{\testRecallOpenai}{97\%}

\newcommand{\testRecallOpenaiFrac}{65/67}

\newcommand{\testRecallPro}{78\%}

\newcommand{\testRecallProFrac}{35/45}

\newcommand{\testUncertain}{13}
\newcommand{\topN}{30}

\newcommand{\verFlag}{23\%}

\newcommand{\verFlagFrac}{117/500}
\newcommand{\verFlagK}{117}

\newcommand{\verOneJudge}{87}
\newcommand{\verResolveFlagged}{31\%}
\newcommand{\verResolveUnflagged}{88\%}
\newcommand{\verScoreFour}{39}
\newcommand{\verScoreThree}{78}
\newcommand{\verTasks}{500}

\newcommand{\code}[1]{\texttt{#1}}
\newcommand{\fit}[1]{\resizebox{\ifdim\width>\linewidth\linewidth\else\width\fi}{!}{#1}}
\title{Spec-Test Coverage Gap: Detecting Overspecified Tests in SWE-Bench-Style Benchmarks}
\author{%
  Avi Aggarwal \\
  Purdue University \\
  \texttt{aggar138@purdue.edu} \\
  \And
  Yiran Hu \\
  Purdue University \\
  \texttt{hu954@purdue.edu} \\
}

\begin{document}
\maketitle

\begin{abstract}
SWE-bench-style benchmarks, such as SWE-bench Verified (Verified) and SWE-bench Pro (Pro), evaluate coding agents with tests taken from the change that originally fixed each issue. However, these tests can enforce details the task never states, causing correct alternative fixes to fail. Such overspecified tests have so far been found mainly through expensive expert review. To address this, we define the Spec-Test Coverage Gap (STCG), a 1--4 scale measuring overspecification, and score it with two LLM judges, Claude Opus 5.5 and GPT-6.1 Sol, which are given the specification and the test patch. A task is \emph{flagged} as overspecified when both judges score it 3 or higher. Our study has three parts. First, we validate the judges on \nTest{} tasks that earlier human audits and agent results label as overspecified or fine. Second, we apply STCG to all \verTasks{} Verified and \proTasks{} public Pro tasks and compare the flags with the results of the top \topN{} Verified submissions. Third, we verify the flags by asking Claude Opus 5.5 to loosen the overspecified requirements of flagged tasks, and regrade \flagCands{} failed patches from those submissions. Our main findings are: (1) Overspecified tests are detectable: on the validation tasks, the judges correctly flag \testRecall{} of the tasks that human audits identify as overspecified and wrongly flag only \testFFR{} of the tasks known to be fine. (2) Overspecified tests are common: applied to whole benchmarks, the judges flag \verFlag{} of Verified and \proFlag{} of Pro tasks. (3) Flagged tasks are much harder for agents: the top \topN{} Verified submissions resolve only \verResolveFlagged{} of the flagged tasks on average, against \verResolveUnflagged{} of the other tasks. (4) Overspecified tests change the leaderboard: relaxing the tests lets \flagFlipAll{} of the \flagCands{} failed patches on flagged tasks pass. Counting these patches as resolved would add \flagGainMin{} to \flagGainMax{} tasks to each top-\topN{} submission, raising resolve rates by up to \flagGainMaxPts{} percentage points. Code and data are available at \url{https://github.com/ObviAvi/Spec-Test-Coverage-Gap}.
\end{abstract}

\section{Introduction}

Coding agents are commonly compared on SWE-bench-style benchmarks, such as SWE-bench \cite{jimenez2024swebench}, SWE-bench Verified \cite{openai2024verified} and SWE-bench Pro \cite{swebenchpro}. Each task gives the agent a code repository and a description of a problem, usually a GitHub issue, and grades the agent's code change (its \emph{patch}) by running tests the agent never sees. The tests come from the pull request or commit that fixed the issue, and most were added or edited to confirm that one fix works, not to accept every fix that works. When a test checks a detail the specification leaves open, such as the exact wording of an error message, a sentinel return value, or the behavior of a private helper, an agent can resolve the reported problem and still be graded as failing. We call such tests \emph{overspecified}. Table~\ref{tab:examples} shows two examples from Verified.

\begin{table}[t]
\caption{Examples of overspecified tasks.}
\label{tab:examples}
\centering\small
\begin{tabular}{@{}p{0.15\linewidth}>{\raggedright\arraybackslash}p{0.24\linewidth}>{\raggedright\arraybackslash}p{0.22\linewidth}>{\raggedright\arraybackslash}p{0.30\linewidth}@{}}
\toprule
Task & The issue asks for & The evaluation test requires & Why it is overspecified \\
\midrule
\code{django-16667} & \code{SelectDateWidget} should not crash with \code{OverflowError} when a request supplies a very large year, month or day. & \code{value\_from\_\allowbreak datadict} returns exactly the string \code{"0-0-0"} for such input. & The issue never says what to return. A fix that returns the submitted values, as the widget already does for other invalid dates, stops the crash but fails the test. \\
\addlinespace
\code{django-13794} & The \code{add} template filter should concatenate a string with a lazy string instead of returning an empty string. & Two lazy \emph{integer} proxies can be added with \code{+}. & The issue is only about the filter. A fix inside the filter passes the filter tests but fails this one, which needs \code{+} support on Django's lazy objects. \\
\bottomrule
\end{tabular}
\end{table}

Test-based grading can be wrong in two directions. Tests that are too weak accept wrong patches and inflate scores. Prior work detects them by generating additional or adversarial tests that reject wrong patches \cite{utboost,sweabs}, or tests that expose behavior differences between an agent's patch and the reference fix \cite{patchdiff}. Tests that are too strict reject correct patches and deflate scores. Human audits suggest this is common among the hardest tasks, the ones top agents still fail: OpenAI reports that at least 59.4\% of the 138 Verified tasks its o3 model does not consistently solve have tests that reject functionally correct submissions \cite{openai2026why}, and in a later audit estimates that about 30\% of the public Pro tasks are broken, with overly strict tests among the main causes \cite{openai2026signal}. Yet strict tests have been found mainly by expert review, and existing automated auditors treat them as one category among many (Section~\ref{sec:related}). Three challenges set them apart. (1) \emph{No executable proof:} a wrong patch that passes is enough to expose a weak test, but a failing patch proves nothing on its own, since it may be wrong or may be correct and rejected. Telling the two apart means judging it against an informally worded specification. (2) \emph{Scarce labels:} expert review is costly to repeat for every task and every new benchmark, so few labeled tasks exist to validate a detector against. (3) \emph{Repairs can weaken tests:} loosening a strict test to accept other fixes can also let wrong patches through, so any repair must be checked in both directions.

This paper asks whether overspecified tests can be detected from the task alone, cheaply and with a measured error rate. We make five contributions.
\begin{itemize}
\item \textbf{A definition and a judge.} The Spec-Test Coverage Gap (STCG) rates a task from 1 (fine) to 4 (significantly overspecified; Table~\ref{tab:scale}). Two LLM judges from different model families rate each task from its specification and test patch only, and the two-judge \emph{panel} flags a task when both rate it 3 or 4 (Section~\ref{sec:method}).
\item \textbf{Validation against existing evidence.} We build validation sets from human annotations and public agent results, with no new manual labeling. On a held-out split the panel flags \testRecall{} of known-overspecified tasks and \testFFR{} of known-fine tasks (Section~\ref{sec:validation}).
\item \textbf{Prevalence.} The panel flags \verFlag{} (\verFlagFrac) of SWE-bench Verified and \proFlag{} (\proFlagFrac) of SWE-bench Pro. On Verified, flagged tasks are resolved by \verResolveFlagged{} of the top \topN{} public submissions on average, against \verResolveUnflagged{} for unflagged tasks (Section~\ref{sec:prevalence}).
\item \textbf{Ablations.} We measure run-to-run noise, the effect of showing the reference fix, and how much of the result depends on the prompt (Section~\ref{sec:ablations}).
\item \textbf{Executable confirmation.} An LLM relaxes the flagged tests, deterministic checks validate each relaxed test, and existing agent patches are regraded, with no new agent runs. On flagged tasks \flagFlipAll{} of the patches that originally fail only the evaluation tests pass the relaxed tests. On tasks that neither judge flags the relaxer leaves all the tests unchanged when allowed to, and when forced to loosen them only \forcedFlipAll{} of such patches pass (Section~\ref{sec:exec}).
\end{itemize}

\section{Background and related work}
\label{sec:related}

\paragraph{Human audits.}
Most of what is known about overspecified tests comes from manual audits. OpenAI builds SWE-bench Verified from 1,699 SWE-bench tasks. Three annotators rate each task on several criteria, including whether the tests may reject valid solutions, and the 500 tasks without major problems are kept \cite{openai2024verified}. Later audits find that problems remain in both benchmarks. On Verified, OpenAI reviews the 138 tasks that o3 does not consistently solve and finds tests that reject correct submissions in at least 59.4\% of them \cite{openai2026why}. On the 731 public Pro tasks, an automated pipeline judges 27.4\% broken and human annotators 34.1\%, which OpenAI summarizes as about 30\% \cite{openai2026signal}. SWE-Bench Pro Verified collects 119 publicly reported problem tasks among the same 731 and revises 102 of them, 75 because the tests are overly narrow: they enforce strings, types, orderings or boundary behavior the task does not specify \cite{proverified}. These audits show that overspecified tests survive human screening; STCG aims to detect them automatically and with a measured error rate.

\paragraph{LLM auditors.}
Several tools use LLMs to audit benchmark tasks. SPICE \cite{spice} asks an LLM to reproduce the annotation scales used for Verified. Against the annotators' majority labels on 110 tasks, it reaches 87.3\% accuracy on issue clarity and 68.5\% on test coverage. Its test-coverage step reads the reference fix and a map of the repository, and its label puts tests that are too narrow and tests that are too broad on one scale. PAIChecker \cite{paichecker} finds that in 13.6\% of Verified tasks the pull request does not match its issue. One of its five patterns, \emph{Unspecified Literal}, is closest to our target and covers tests that assert exact values the issue never mentions. Auto Benchmark Audit \cite{aba} is an agent that audits 168 benchmarks, rating each task on its instructions, environment and evaluation. Tests that reject valid alternatives are one of several evaluation failures. Task Verification Bench \cite{taskverificationbench} scores auditors against defects that benchmark maintainers have acknowledged, and finds that counting a task as detected whenever an auditor flags it overstates recall by up to 54 percentage points. STCG differs in three ways: it targets one defect, reads only what the agent is told plus the tests, and is validated against labels for that defect.

\paragraph{Weak tests.}
A separate line of work studies the opposite problem: tests too weak to reject wrong patches. UTBoost \cite{utboost} adds LLM-generated test cases to the existing tests. SWE-ABS \cite{sweabs} adds tests that reach untested code, then generates plausible but wrong patches and writes tests that reject them. On the top 30 Verified submissions, its stronger tests reject about 20\% of the patches the original tests accepted, and all 30 submissions change rank. PatchDiff \cite{patchdiff} generates tests that expose behavior differences between an agent's patch and the reference fix. Our work is the mirror image: tests that reject correct patches.

\section{The Spec-Test Coverage Gap}
\label{sec:method}

\subsection{Definition}

STCG measures how far a task's tests go beyond what the agent is told. A task consists of a specification, which is everything the agent is told, and two sets of tests. The \emph{evaluation tests} must fail before the fix and pass after it (FAIL\_TO\_PASS in SWE-bench), and STCG rates only these. The \emph{regression tests} must pass both before and after (PASS\_TO\_PASS) and check that existing behavior still works. A patch resolves the task only if it passes both. STCG is based on the following question:

\begin{quote}
Picture competent developers who work in this repository, have read the specification, and have seen neither the tests nor the original fix. Each writes a correct fix. Would every such fix pass the evaluation tests?
\end{quote}

The judge answers this question with a score from 1 to 4 (Table~\ref{tab:scale}). Only reasonable fixes count: a fix that a competent developer could plausibly write from the specification, that resolves the reported case and closely related inputs. Scores 1 and 2 mean every reasonable correct fix passes; scores 3 and 4 mean some reasonable correct fix fails (Table~\ref{tab:examples} gives two examples). All validation in this paper uses the binary reading: a task is \emph{fine} (1--2) or \emph{overspecified} (3--4). We retain the distinction between scores 3 and 4 because the severity of overspecification may affect how a task should be repaired.

\begin{table}[t]
\caption{The STCG scale. A task is flagged when its final score is 3 or 4.}
\label{tab:scale}
\centering\small
\begin{tabular}{@{}cp{0.88\linewidth}@{}}
\toprule
Score & Meaning \\
\midrule
1 & \textit{Directly specified.} The evaluation tests check the behavior the specification describes, or its reproduction example, near-verbatim. \\
\addlinespace
2 & \textit{Reasonably specified.} The tests add only what a careful reader would infer: a standard exception type, an existing pattern, the same function on related inputs. \\
\addlinespace
3 & \textit{Mildly overspecified.} The tests check observable behavior but pin one valid choice among several, such as an exact message, an order, or a specific value. \\
\addlinespace
4 & \textit{Significantly overspecified.} The tests check how the fix was written: a sentinel value, internal state, a private method, a new name the specification never mentions, or a different code path than the one the specification implicates. \\
\bottomrule
\end{tabular}
\end{table}

\subsection{The judge}

\paragraph{Input.}
The judge sees only what an agent sees, plus the tests. The prompt lists the evaluation tests by name and, for Python, also marks each one where it is defined in the patch. For SWE-bench the specification is the issue text. For SWE-bench Pro it also includes the \code{requirements} and \code{interface} fields, since agents receive them. The judge does not see the repository or the reference fix. This keeps the judgment grounded in the information provided to an agent during an actual attempt, requires only one model call per task per judge, and avoids anchoring the assessment on the reference fix.
\paragraph{Prompt.}
The prompt states the question above, defines what constitutes a reasonable fix, and provides the four-point scale with examples and tie-breaking rules for adjacent scores. It then asks the judge to enumerate the requirements imposed by the evaluation tests, classify each as specified, established by convention or existing code, or left open, and describe a reasonable correct fix that would violate each open requirement. The full text is in Appendix~\ref{app:prompt}. The judge returns structured output with four fields: its reasoning, the score, a one- or two-sentence explanation naming what a correct fix would get wrong, and the test lines responsible.

\paragraph{Panel and aggregation.}
To reduce false flags, two models from different providers judge each task with the same prompt: \code{claude-opus-5-5} and \code{gpt-6.1-sol}, both at high reasoning effort. The final score is the lower of the two, so a task is flagged only when both judges flag it. Tasks whose scores differ by two or more points are marked \emph{uncertain} and counted separately.

\section{Validation}
\label{sec:validation}

\subsection{Validation sets}

Measuring the judge's accuracy requires tasks whose answer is already known. We build such tasks from existing human annotations and public agent results, without new manual labeling (Table~\ref{tab:sets}).

\begin{table}[t]
\caption{Validation sets. The split is by task, stratified by set, with 40\% in dev.}
\label{tab:sets}
\centering\small
\fit{\begin{tabular}{llp{6.1cm}rrr}
\toprule
Set & Label & Source & Tasks & Dev & Test \\
\midrule
\texttt{openai\_excluded} & overspecified & SWE-bench tasks left out of Verified; all three annotators said the tests may reject valid fixes & 112 & 45 & 67 \\
\addlinespace
\texttt{pro\_narrow\_original} & overspecified & SWE-bench Pro tasks later repaired for overly narrow tests (original versions) & 75 & 30 & 45 \\
\addlinespace
\texttt{agent\_consensus} & fine & Verified tasks all three annotators rated fine, resolved by $\geq$90\% of top submissions with differing patches & 157 & 63 & 94 \\
\midrule
Total & & & 344 & 138 & 206 \\
\bottomrule
\end{tabular}
}
\end{table}

\paragraph{Known overspecified.}
Positive examples come from two human audits. \code{openai\_excluded} contains the SWE-bench tasks that OpenAI leaves out of Verified for which all three annotators rate the tests as likely to reject valid solutions, rate the issue itself as clear, and record no other major problem \cite{openai2024verified}. \code{pro\_narrow\_original} contains the 75 public SWE-bench Pro tasks that SWE-Bench Pro Verified repairs for overly narrow tests \cite{proverified}, in their original, unrepaired versions.

\paragraph{Known fine.}
Negative examples are tasks that both annotators and agents treat as fine. \code{agent\_consensus} contains Verified tasks where all three annotators find no problem with the tests, at least 90\% of the top \topN{} public submissions (ranked by tasks resolved) resolve the task \cite{swebenchexperiments}, and their resolving patches differ from one another (mean pairwise Jaccard similarity of changed lines at most 0.5), so agents pass with different fixes rather than one shared fix. We find no comparable evidence for Pro, since the repaired Pro tasks may still contain other overspecified checks, so the false-flag rate is measured on Verified only.

\subsection{Protocol}

We develop the prompt on one part of the data and report results on another, so that tuning cannot inflate the reported accuracy. The prompt is developed on the dev split (Table~\ref{tab:sets}) and then frozen, and the test split is scored once with the frozen prompt. We measure \emph{recall}, the share of known-overspecified tasks the panel flags, and the \emph{false-flag rate}, the share of known-fine tasks it flags. Targets are fixed in advance: recall of at least 70\% and a false-flag rate of at most 15\%. We report recall and false-flag rate with Wilson 95\% intervals \cite{wilson1927}, and agreement between the judges as Cohen's $\kappa$ on the binary flag \cite{cohen1960}. All \nValidation{} tasks are judged by both models with no errors.

\subsection{Results}

\begin{table}[t]
\caption{Detection on the validation sets. Parentheses give counts and 95\% CIs.}
\label{tab:validation}
\centering\small
\fit{\begin{tabular}{lllll}
\toprule
Scorer & Recall, all overspecified & \texttt{openai\_excluded} & \texttt{pro\_narrow\_original} & False-flag rate \\
\midrule
\multicolumn{5}{l}{\textit{Dev split} (138 tasks)} \\
\quad Panel (both judges flag) & 81\% (61/75; 71--89\%) & 87\% (39/45) & 73\% (22/30) & 3\% (2/63; 1--11\%) \\
\quad Claude alone & 83\% (62/75; 73--90\%) & 89\% (40/45) & 73\% (22/30) & 3\% (2/63; 1--11\%) \\
\quad GPT alone & 88\% (66/75; 79--94\%) & 89\% (40/45) & 87\% (26/30) & 17\% (11/63; 10--29\%) \\
\midrule
\multicolumn{5}{l}{\textit{Test split} (206 tasks)} \\
\quad Panel (both judges flag) & 89\% (100/112; 82--94\%) & 97\% (65/67) & 78\% (35/45) & 4\% (4/94; 2--10\%) \\
\quad Claude alone & 89\% (100/112; 82--94\%) & 97\% (65/67) & 78\% (35/45) & 4\% (4/94; 2--10\%) \\
\quad GPT alone & 94\% (105/112; 88--97\%) & 97\% (65/67) & 89\% (40/45) & 18\% (17/94; 12--27\%) \\
\bottomrule
\end{tabular}
}
\end{table}

On the test split (Table~\ref{tab:validation}), the panel flags \testRecall{} of the known-overspecified tasks and \testFFR{} of the known-fine tasks. Both targets are met on both splits. The judges agree on the flag for \testAgree{} of test tasks ($\kappa = \testKappa$), and \testUncertain{} test tasks are uncertain.

\paragraph{Recall differs by source.}
On the test split the panel flags \testRecallOpenai{} (\testRecallOpenaiFrac) of \code{openai\_excluded} but \testRecallPro{} (\testRecallProFrac) of \code{pro\_narrow\_original}. Pro specifications are longer and state names, signatures and messages in their \code{requirements} and \code{interface} fields, so more of what a test checks counts as specified.
\paragraph{What the second judge adds.}
Requiring both judges to agree removes most false flags at little cost in recall. GPT alone has higher recall than the panel (\testGptRecall{} on test) but flags \testGptFFR{} of the fine tasks, while Claude alone matches the panel on test in both rates.

\paragraph{False flags.}
The \testFalseFlags{} labeled-fine tasks flagged by the panel are cases where both judges identify a specific implementation choice that the tests require but the specification leaves open. Examples include a fixed ordering for a de-duplicated list, an exact error message, an array shape, and a particular internal collection. Nearly all top submissions pass these tasks, suggesting that agents generally make the choices expected by the tests.
\section{Prevalence}
\label{sec:prevalence}
\suppressfloats[t]  

To estimate how common overspecified tests are, we run the frozen judges on all 500 SWE-bench Verified tasks and all 731 public SWE-bench Pro tasks (Table~\ref{tab:prevalence}).

\begin{table}[t]
\caption{Flag rates on SWE-bench Verified and the public SWE-bench Pro tasks.}
\label{tab:prevalence}
\centering\small
\fit{\begin{tabular}{lll}
\toprule
& Verified & Pro (public) \\
\midrule
Tasks & 500 & 731 \\
Flagged by the panel & 23\% (117/500) & 35\% (258/731) \\
\quad 95\% CI & 20--27\% & 32--39\% \\
\quad score 3 (mild) & 78 & 143 \\
\quad score 4 (significant) & 39 & 115 \\
$\kappa$ between judges & 0.61 & 0.62 \\
Uncertain (spread $\geq$2) & 11\% (57/500) & 10\% (71/731) \\
\bottomrule
\end{tabular}
}
\end{table}

\paragraph{Verified.}
The panel flags \verFlag{} of Verified: \verScoreThree{} tasks at score 3 and \verScoreFour{} at score 4. These tasks have already passed a human screening process intended to exclude problems of this kind, yet the panel still flags a substantial fraction of them. Among the seven repositories with at least 20 tasks the rate ranges from 20\% to 31\% (Appendix~\ref{app:repos}).

\paragraph{Pro.}
The panel flags \proFlag{} of the public Pro tasks. The \nProNarrow{} validation tasks are part of this set and are known to be overspecified. Among the remaining tasks the rate is \proFlagRest{} (\proFlagRestFrac). A larger share of flagged Pro tasks receive score 4 than on Verified (\proScoreFour{} of \proFlagK{} against \verScoreFour{} of \verFlagK).

\subsection{Flags and agent results on Verified}
\label{sec:resolve}

If flagged tasks reject correct fixes, agents should resolve them less often. We compare flags with the per-task results of \nSubmissions{} public Verified submissions \cite{swebenchexperiments}.

Agents fail flagged tasks far more often: flagged tasks are resolved by \verResolveFlagged{} of the top \topN{} submissions on average, unflagged tasks by \verResolveUnflagged. The flag rate falls as more submissions resolve a task, from \bandZero{} of tasks that no top submission resolves to \bandHigh{} of tasks that at least 90\% resolve (Figure~\ref{fig:resolve} in Appendix~\ref{app:resolve}).
However, difficulty alone does not fully explain the gap. Harder tasks are flagged more often, but at every difficulty level flagged tasks have a mean resolve rate more than 50 percentage points below unflagged tasks (Appendix~\ref{app:difficulty}).

A lower resolve rate is expected for overspecified tests, but it does not necessarily show that the failing patches are correct. The judge could also be picking up other properties of harder tasks. Section~\ref{sec:exec} tests this directly by relaxing the flagged tests and regrading the failing patches.

\section{Ablations}
\label{sec:ablations}
\suppressfloats[t]  

This section checks how much the results depend on three choices: random variation between runs, whether the judge sees the reference fix, and how detailed the prompt is. Table~\ref{tab:ablation} compares each variant with the main prompt on the dev split. \emph{Flags changed} counts tasks whose panel flag differs from the first run of the full prompt.

\begin{table}[t]
\caption{Ablations on the dev split (\nAblation{} tasks).}
\label{tab:ablation}
\centering\small
\fit{\begin{tabular}{lllrrrrr}
\toprule
Prompt & Recall & False flags & \multicolumn{2}{c}{Recall, one judge} & $\kappa$ & Uncertain & Flags \\
 & & & Claude & GPT & & & changed \\
\midrule
Full prompt & 81\% (61/75) & 3\% (2/63) & 83\% & 88\% & 0.78 & 10 & -- \\
Full prompt, identical rerun & 80\% (60/75) & 5\% (3/63) & 81\% & 88\% & 0.77 & 12 & 6 \\
Full prompt + reference fix & 87\% (65/75) & 5\% (3/63) & 87\% & 88\% & 0.87 & 5 & 5 \\
One-paragraph baseline & 72\% (54/75) & 6\% (4/63) & 81\% & 77\% & 0.78 & 4 & 9 \\
\bottomrule
\end{tabular}
}
\end{table}

\paragraph{Run-to-run noise.}
Running the identical prompt a second time changes the panel flag on \rerunChanged{} of \nAblation{} tasks, in both directions; recall and the false-flag rate each change by one task. Differences of one or two tasks between prompts are therefore comparable to run-to-run variation.

\paragraph{Showing the reference fix.}
The reference fix shows the judge one correct solution, which could help it or anchor it on that one solution. Adding it changes \goldChanged{} panel flags, about as many as the rerun. Recall (\goldRecall{} against \fullRecall) and the false-flag rate (\goldFFR{} against \fullFFR) do not differ measurably on \nAblation{} tasks. The clearer effect is on agreement: $\kappa$ rises from \fullKappa{} to \goldKappa{} and uncertain tasks fall from \fullUncertain{} to \goldUncertain. Since the judge does about as well without it, we keep the reference fix out of the input.

\paragraph{A one-paragraph prompt.}
To see how much the detailed prompt matters, we replace its instructions with one paragraph that names the four scores without defining them, keeping the inputs, models and output fields (Appendix~\ref{app:baseline}). Recall falls from \fullRecall{} to \baselineRecall{} and the false-flag rate rises from \fullFFR{} to \baselineFFR; the baseline loses \baselineLost{} overspecified flags and gains none, against \rerunLost{} lost and \rerunGained{} gained in the rerun. The loss comes almost entirely from GPT (recall \fullGptRecall{} to \baselineGptRecall, while Claude's is nearly unchanged at \fullClaudeRecall{} to \baselineClaudeRecall) and from Pro tasks (\fullRecallProFrac{} to \baselineRecallProFrac). Exact scores also drift: each judge gives the same 1--4 score as under the full prompt on about two thirds of tasks, against about 90\% in the rerun.

Even the one-paragraph prompt meets both validation targets; the detailed prompt adds recall for one judge on Pro and stabilizes the four-point scale. Our contribution is the measured accuracy of the pipeline, not the prompt itself.

\section{Executable confirmation}
\label{sec:exec}
\suppressfloats[t]  

The judges only read text, so we also test their flags by executing the test code. A flag means that some reasonable correct fix would fail the evaluation tests. Public submissions may already contain such fixes among their failing patches. If a flagged task's tests are overspecified in this way, then loosening that requirement, and nothing else, should let some of those patches pass. We relax the tests of flagged tasks and regrade existing patches, without rerunning any agent.

\subsection{Method}

\paragraph{Candidate patches.}
We need failed patches that might nevertheless be correct, so we get them from the patches published by the top \topN{} Verified submissions \cite{swebenchexperiments}. A \emph{candidate} is a failed patch whose report shows at least one failing evaluation test and no failing regression test. We skip patches that fail regression tests, since they have problems beyond the requirements being evaluated. On flagged tasks, \flagNewOnly{} of the \flagFailing{} failing patches fail only evaluation tests, against \ctrlNewOnly{} of \ctrlFailing{} on tasks neither judge flags. Of the \flagGroup{} flagged tasks, \flagTasks{} have at least one candidate, with \flagCands{} candidates in total.

\paragraph{Relaxer.}
The \emph{relaxer} is an LLM (\code{claude-opus-5-5}) that edits a flagged task's tests so they accept other reasonable fixes. It receives the specification, the evaluation test names, the test patch, the reference fix, and both judges' explanations with the test lines they point to. It is told to loosen only requirements the specification leaves open, keep every other requirement, and decline if it disagrees with the judges. Its edits must fall inside the test patch, it may not add, rename or delete tests, and it never sees an agent patch (prompt in Appendix~\ref{app:relax}). We develop the prompt on tasks that exactly one judge flags, which are excluded from all results here. It makes \relaxCalls{} calls for the \flagTasks{} flagged tasks, averaging \relaxTokensIn{} input and \relaxTokensOut{} output tokens per call (Appendix~\ref{app:cost}).

\paragraph{Checks.}
Relaxed tests run in the task's official container image with the SWE-bench harness \cite{jimenez2024swebench}. We consider a relaxation \emph{valid} only when (1) at least one evaluation test still fails on the unfixed repository, (2) the reference fix passes every evaluation and regression test, and (3) the patches that resolve the original task still resolve it. These are all resolving top-\topN{} patches, plus resolving patches from lower-ranked submissions when there are fewer than five. After a failed check, the relaxer sees the failure and the test output and may retry, up to three attempts in total.

\paragraph{Regrading.}
We then rerun each candidate patch against the valid relaxed tests. We count a patch as a \emph{flip} when it passes every evaluation and regression test under the relaxed tests. Candidates come from the submitters' published reports, so we also rerun the original tests ourselves. A patch whose published failure does not reproduce in our environment does not count as a flip. Our main measure is the share of all candidate patches that flip. If a task has no valid relaxation, all of its candidates count as not flipping. Patches from one task share the same relaxed tests, so their outcomes are not independent. We therefore compute 95\% confidence intervals by resampling tasks, not patches, with replacement 10,000 times.

\paragraph{Controls.}
A relaxed test is weaker, and a weaker test passes more patches whether or not the original is overspecified. To measure this, we run the same procedure on the \forcedGroup{} Verified tasks that neither judge flags, of which \forcedTasks{} have a candidate (\forcedCands{} candidates). We call these the \emph{control} tasks. The \verOneJudge{} tasks that exactly one judge flags are in neither group, because we use them to develop the relaxer prompt. Leaving out these borderline tasks makes the two groups easier to separate than flagged and unflagged tasks in general, so the comparisons below are between clear cases on both sides. These tasks have no judge findings, so we use two prompts without them. The \emph{forced} prompt says the tests have been reviewed and found fine, but requires loosening the one requirement that a differently written correct fix is most likely to miss. The \emph{no-findings} prompt is the main prompt without the findings, and it lets the relaxer decline. We also run the no-findings prompt on the \nofindTasks{} flagged tasks, so the two groups can be compared under one prompt.

\subsection{Results}

\begin{table}[t]
\caption{Relaxing tests and regrading candidate patches.}
\label{tab:exec}
\centering\small
\fit{\begin{tabular}{lrrrr}
\toprule
& \multicolumn{2}{c}{Both judges flag} & \multicolumn{2}{c}{Neither judge flags} \\
 & with findings & no findings & forced & no findings \\
\midrule
Tasks & 108 & 108 & 79 & 79 \\
Candidate patches & 2,186 & 2,186 & 417 & 417 \\
Valid relaxations & 93 & 82 & 76 & 0 \\
Declined & 15 & 26 & 1 & 79 \\
Lines changed, median & 11 & 10 & 5 & -- \\
\midrule
Flip rate, all candidates & 61\% & 49\% & 29\% & 0\% \\
\quad 95\% CI & 52--69\% & 40--58\% & 13--45\% & -- \\
Flip rate, regraded patches & 72\% & 64\% & 32\% & -- \\
Tasks with no flip & 22/108 & 40/108 & 60/79 & 79/79 \\
\bottomrule
\end{tabular}
}
\end{table}

\paragraph{Flagged tasks, with the judges' findings.}
The relaxer produces a valid relaxation for \flagValid{} of the \flagTasks{} flagged tasks (\flagFirst{} on the first attempt) and declines the other \flagDeclined, answering that the tests match the specification (Table~\ref{tab:exec}). A valid relaxation changes a median of \flagLinesMedian{} lines of test code. Of the total \flagCands{} candidates, \flagFlipAllK{} (\flagFlipAll) flip. Also, on \flagTasksWithFlipK{} of the \flagValid{} relaxed tasks, at least one patch flips.

\paragraph{Forced control.}
The forced control shows how many patches pass when tests that are not flagged are loosened anyway. Under the forced prompt, the relaxer produces a valid relaxation for \forcedValid{} of the \forcedTasks{} control tasks. It declines \forcedDeclined, and \forcedInvalid{} fail the checks. Of the \forcedCands{} candidates, \forcedFlipAllK{} (\forcedFlipAll) flip, against \flagFlipAll{} on flagged tasks. The 95\% CI of the difference is \flipDiffCI{} percentage points. Most control tasks have no flip at all: \forcedNoFlipK{} of \forcedTasks, against \flagNoFlipK{} of \flagTasks{} flagged tasks.

The control flips come mostly from five tasks, which account for \forcedTopFiveFlips{} of the \forcedFlipAllK. On the other \forcedRestTasks{} control tasks, only \forcedRest{} of candidates flip. Both judges score each of the five at 2, one step below a flag. For each, we inspect the issue, the relaxation and two to four flipped patches. In four of the five tasks, the inspected patches fix the case the issue reports but miss a related case that the removed requirement covers. The fifth is borderline. This review suggests that many control flips come from weakening a test that is not overspecified, not from overspecification the panel missed.

\paragraph{One prompt on both groups.}
A stricter comparison gives both groups the same prompt, without the judges' findings. Under this no-findings prompt the relaxer declines all \freeTasks{} control tasks, so no patch flips. On flagged tasks the same prompt relaxes \nofindValid{} of \nofindTasks{} tasks and \nofindFlipAll{} of candidates flip, against none on control tasks (95\% CI of the difference \nofindDiffCI{} percentage points). The relaxer therefore separates them without seeing the judges' findings. Showing the findings raises the flip rate from \nofindFlipAll{} to \flagFlipAll{} but does not create the gap.

Flagged tasks are harder, but within every resolve band the flip rate is higher on flagged tasks (Appendix~\ref{app:bands}). The forced control also changes fewer lines (median \forcedLinesMedian{} against \flagLinesMedian), but on flagged tasks the flip rate does not grow with relaxation size: \flagFlipSmall{}, \flagFlipMid{} and \flagFlipLarge{} for relaxations of 1--5, 6--15 and 16 or more lines.

\paragraph{Effect on scores.}
If every flip counted as resolved, each top-\topN{} submission would gain \flagGainMin{} to \flagGainMax{} of the 500 tasks (median \flagGainMedian) and \flagRankChanged{} of the \topN{} would change rank, by up to \flagRankMove{} places (Appendix~\ref{app:agents}). Under the forced control, the gain is \forcedGainMin{} to \forcedGainMax{} tasks. These gains are an upper bound on the credit overspecified tests withhold, not corrected scores (Section~\ref{sec:limits}).

\section{Limitations}
\label{sec:limits}

\begin{itemize}
\item \textbf{Pro is checked less than Verified.} There are no known-fine Pro tasks, so the judge's false-flag rate on Pro is unknown, and the Pro prevalence figures assume it is similar. The executable confirmation also covers Verified only.
\item \textbf{The validation sets contain clear cases.} Positives are tasks that annotators or experts singled out. Negatives are tasks that annotators cleared and nearly every agent resolves, which shows that agents converge on what the tests expect, not that no reasonable alternative exists. Accuracy on borderline tasks is not measured.
\item \textbf{A flip is not a proof.} We inspect only a sample of flipped patches: two to four on each of five control tasks, and the examples in Appendix~\ref{app:flips}. A patch that passes the relaxed tests may be wrong in the part the relaxation removes, so the score changes in Section~\ref{sec:exec} are upper bounds.
\item \textbf{The checks bound a relaxation from one side only.} A relaxed test must still fail the unfixed code and pass the reference fix, but nothing stops it from being looser than the specification warrants. The pylint relaxation in Appendix~\ref{app:flips}, for example, accepts any absolute path that names pylint. A correct fix passes it, but so would a wrong one such as \code{/tmp/pylint}, which the issue does not support.
\item \textbf{The relaxer is not independent of the judges.} It is the same model as one judge, and in the main run it reads both judges' findings. When it relaxes most flagged tasks and declines every control task under one prompt, it partly repeats a verdict that model already gave.
\item \textbf{No repository context.} The judge cannot see existing conventions that would settle a detail, or existing code that would make a test's requirement unreachable. This can cause both false flags and misses.
\item \textbf{No comparison with other auditors.} We do not run SPICE or other auditors on the same tasks.
\end{itemize}

\section{Conclusion}

Overspecified tests can be detected from a task's specification and tests alone, with a measured error rate. First, on the test split the two-judge panel flags \testRecall{} of known-overspecified tasks and only \testFFR{} of known-fine ones. Second, the panel flags about a quarter of SWE-bench Verified and a third of public SWE-bench Pro tasks as having evaluation tests that go beyond what the agent is told. On Verified these tasks are concentrated among those that top agents fail. Third, running the tests supports the flags. Once the flagged tests are relaxed and the relaxed tests pass our checks, \flagFlipAll{} of the patches that originally fail only the evaluation tests pass. On tasks that neither judge flags, the relaxer leaves every test unchanged when it is allowed to, and when it is forced to loosen them, only \forcedFlipAll{} of such patches pass.

We see three directions for future work. The first is the other repair: adding the missing requirement to the specification instead of loosening the test, and measuring how much of the gap agents close when they are told what the tests expect. The second is contamination. The tests of a flagged task require details the issue does not state, so an agent that passes them has guessed those details, inferred them from the repository, or recalled them from training. Asking models to reproduce these details from the issue alone, with and without the repository, and comparing the results across model release dates would help separate recall from inference. The third is coverage: confirming flags by execution on Pro, running the panel on other benchmarks, and comparing it with SPICE on the same tasks.

\clearpage
\bibliographystyle{unsrtnat}
\bibliography{refs}

\clearpage
\appendix

\section{Judge prompt}
\label{app:prompt}

The frozen prompt \code{judge\_v1}. The three placeholders in double braces are replaced with the task's content. The prompt calls the evaluation tests the \emph{graded} tests.

\begin{Verbatim}
You are auditing one task from a software-engineering benchmark such as SWE-bench.

In these benchmarks an agent receives a task specification (a bug report or feature request) and the repository, writes a fix, and is graded by tests it never sees. The tests come from the pull request that originally fixed the issue, so they were written to confirm that one fix. Sometimes they check details the specification never settles, and then a different but correct fix is graded as a failure. Decide whether this task's tests do that, and how badly.

<specification>
{{SPEC}}
</specification>

<fail_to_pass_tests>
{{FAIL_TO_PASS}}
</fail_to_pass_tests>

<test_patch>
{{TEST_PATCH}}
</test_patch>

How to read the inputs:
- The specification is everything the developer is told about the task. Anything it states explicitly counts as specified, including names, signatures, messages, and values given in a Requirements or Interface section.
- The FAIL_TO_PASS tests are the graded ones: they fail before the fix and must pass after it. Python test definitions that are graded are also tagged `# <-- FAIL_TO_PASS` in the test patch. Tests that are not graded do not matter.
- In the test patch, `+` lines are added test code, `-` lines are removed test code, and lines starting with a space are unchanged context. Score what the `+` lines require. The `-` and context lines show how the code already behaved and what already exists.
- You cannot see the repository. The specification and the `-` and context lines are your evidence of what already exists in it.

The question:

Picture competent developers who work in this repository, have read the specification, and have seen neither these tests nor the original fix. Each writes a correct fix: one that resolves what the specification describes, for the reported case and for closely related inputs. Would every such fix pass the graded tests?

Only reasonable fixes count. A fix counts if a competent developer could plausibly write it from this specification. A contrived fix, a fix that ignores part of the specification, or a fix that patches only the literal example and leaves the same bug in closely related inputs does not count.

A detail the tests check is settled if the specification states it or leaves only one reasonable choice. Mentioning a function, field, or behavior does not settle every value, order, name, or edge-case result a test checks on it.

Score from 1 to 4.

Every reasonable correct fix passes:

1 - Directly specified. The graded tests check the behavior the specification describes, or its reproduction example, near-verbatim. No outside assumptions, private APIs, or arbitrary choices.
  Example: the issue says "`q.clear()` leaves items behind in the queue." The test is `q.clear(); assert len(q) == 0`.

2 - Reasonably specified. The tests add only what a careful reader would infer: a standard exception type, behavior that matches an existing pattern, basic null or regression checks, or the same function run on related inputs. What they assert is the only reasonable reading of the specification.
  Example: the issue says "`round_half(2.5)` returns 2.0 instead of 3.0." The test checks `round_half(2.5) == 3.0` and also `round_half(-2.5) == -3.0`.

Some reasonable correct fix fails:

3 - Mildly overspecified. The tests check externally observable behavior but pin one valid choice among several, such as an exact message, an order, or a specific value. A correct fix that made a different reasonable choice fails.
  Example: the issue says "Exporting with an unknown format should give a clear error instead of a KeyError." The test requires the exact message `"Unsupported format: 'xls'"`, so a fix that raises a clear error with different wording fails.

4 - Significantly overspecified. The tests check how the fix was written, not whether the problem is gone. They require a magic or sentinel value, assert internal or intermediate state, call a private method directly, depend on a name the fix must introduce that the specification never mentions, or exercise a different code path than the one the specification implicates. A correct fix made through the natural entry point can fail.
  Example: the issue says "Saving a user whose name contains an emoji crashes." The test skips `save()`, calls the private `_normalize_name()` directly, and asserts the emoji is replaced with `"?"`. Or the issue shows the bug through `Report.render()` but the test calls `Formatter.format_cell()` directly and checks its return value.

Scoring rules:
- Score the most overspecified graded assertion. One assertion that rejects a reasonable correct fix is enough for a 3 or 4.
- Score what a correct fix would actually fail on. A private method or specific value that only appears in the test is not a problem unless a graded assertion depends on it.
- 2 or 3: is the asserted detail the only reasonable choice (2), or one of several the author happened to pick (3)?
- 3 or 4: do the tests assert observable external behavior (3), or internal state, a sentinel, a private method, an unspecified new name, or a different code path (4)?
- Different arguments to the same function are the same code path. A different function is a different code path.
- If you can describe a reasonable correct fix that fails, score 3 or 4, even if you expect most developers would make the same choice the tests did. The question is whether a reasonable alternative is rejected, not which fix is most likely.

Work through these steps before scoring:

1. What does the specification say is wrong or missing? Does it name a function or entry point, or only describe a symptom?
2. List what the graded tests require of a fix: each function, class, constant, or field they call or reference; each value, order, type, and message they assert; each input they use; each mock or patched attribute the fix must read, call, or define.
3. Mark each requirement as settled by the specification (quote the text), settled by convention or existing code (say which), or open.
4. For each open requirement, describe a reasonable correct fix that would violate it, or say why no reasonable fix would.
5. Choose the score.

Return JSON with these fields:
- `reasoning`: a compact pass through the steps: the requirements, each marked settled (with the specification text), convention, or open; and for each open one, the reasonable fix that violates it or why there is none.
- `score`: 1, 2, 3, or 4.
- `explanation`: for a score of 3 or 4, one or two sentences naming what a reasonable correct fix would get wrong against the tests. A later repair step reads this, so be concrete. For a score of 1 or 2, one sentence on why every reasonable fix passes, naming the assertion that maps to the specification.
- `overspecification_vector`: the exact line(s), value(s), or method(s) in the test patch that cause the gap, or "None" for a score of 1 or 2.
\end{Verbatim}

\section{Baseline prompt}
\label{app:baseline}

\begin{Verbatim}
Below is one task from a software-engineering benchmark: the specification the developer is given, the names of the tests that grade the fix (FAIL_TO_PASS), and the test patch that adds them. Decide whether the graded tests are overspecified, meaning they would reject a correct fix that differs from the original one. Score from 1 to 4: 1 = directly specified, 2 = reasonably specified, 3 = mildly overspecified, 4 = significantly overspecified. Return JSON with `reasoning`, `score`, `explanation` (one or two sentences on why), and `overspecification_vector` (the test lines responsible, or "None").

<specification>
{{SPEC}}
</specification>

<fail_to_pass_tests>
{{FAIL_TO_PASS}}
</fail_to_pass_tests>

<test_patch>
{{TEST_PATCH}}
</test_patch>
\end{Verbatim}

\section{Relaxer prompt}
\label{app:relax}

The prompt \code{relax\_v2}, used on flagged tasks. The five placeholders in double braces are filled with the task's content, and, as in the judge prompt, ``graded tests'' means the evaluation tests. Both control prompts drop the judges' findings. The \emph{no-findings} prompt keeps the rest of the text and first asks whether the tests need any change. The \emph{forced} prompt instead says the tests have been reviewed and found fine, and requires loosening exactly one requirement.

\begin{Verbatim}
You are repairing the tests of one task from a software-engineering benchmark such as SWE-bench.

In these benchmarks an agent receives a task specification (a bug report or feature request) and the repository, writes a fix, and is graded by tests it never sees. The tests come from the pull request that originally fixed the issue, so they were written to confirm that one fix. Reviewers found that this task's tests check details the specification never settles, so a different but correct fix can be graded as a failure. Change the tests as little as possible so that they accept every correct fix and still reject the unfixed code.

<specification>
{{SPEC}}
</specification>

<fail_to_pass_tests>
{{FAIL_TO_PASS}}
</fail_to_pass_tests>

<test_patch>
{{TEST_PATCH}}
</test_patch>

<original_fix>
{{GOLD_PATCH}}
</original_fix>

<reviewer_findings>
{{FINDINGS}}
</reviewer_findings>

How to read the inputs:
- The specification is everything the developer is told about the task. Anything it states explicitly counts as specified.
- The FAIL_TO_PASS tests are the graded ones: they fail before the fix and must pass after it. Python test definitions that are graded are also tagged `# <-- FAIL_TO_PASS` in the test patch (the tag is not in the file).
- In the test patch, `+` lines are added test code, `-` lines are removed test code, and lines starting with a space are unchanged context.
- The original fix is one correct fix. It shows which code is involved, and it must pass your tests. It is not the only correct fix: do not write tests that only it would pass.
- Each reviewer finding names what a reasonable correct fix would get wrong against the tests, and the lines responsible. The findings can be wrong or incomplete. Check them against the specification.
- You cannot see the repository or run code.

What to change:
- Loosen only requirements the specification leaves open. Typical repairs: an exact message becomes a check on the exception type or on the part of the message the specification fixes; one pinned value among several valid ones becomes a check that accepts every valid one; a pinned order becomes an order-independent comparison; an assertion on internal state or a private helper becomes a check through the public behavior the specification describes, or is removed if the other assertions already cover that behavior.
- If an entire graded test checks only behavior the specification never asks for, replace its body with `pass`. Keep the test itself so that its name still exists.
- Keep every requirement the specification states or leaves only one reasonable choice for. At least one graded test must still fail on the unfixed code because of the problem the specification reports.
- Do not add requirements. Every fix that passed the original tests must pass yours.
- Touch nothing else. Leave assertions alone when every reasonable correct fix would satisfy them.

Mechanics:
- Give your changes as search-and-replace edits on the test files as they are after the test patch is applied. `old` must be copied exactly from the `+` and context lines of the test patch, without the leading diff character and without the `# <-- FAIL_TO_PASS` tag, with indentation and line breaks intact. It must occur exactly once in the file, so include enough neighboring lines to make it unique. `new` replaces it.
- Edit only inside the regions the test patch shows.
- Do not rename, add, or delete test functions or classes: tests are graded by name, and a skipped or missing test counts as failed.
- Use only names the test patch shows are available in the file, or the standard library. Put a new import inside the test function that needs it.
- If no edit can make the tests accept other correct fixes while still failing on the unfixed code, set `relaxable` to false. This is the right answer when the findings are wrong, or when the only graded test checks nothing the specification asks for.

Return JSON with these fields:
- `relaxable`: true or false.
- `edits`: the list of edits, each with `file` (path as in the test patch), `old`, and `new`. Empty when `relaxable` is false.
- `summary`: one or two sentences on what you loosened, or why nothing can be.
\end{Verbatim}

\section{Flag rate by resolve rate}
\label{app:resolve}

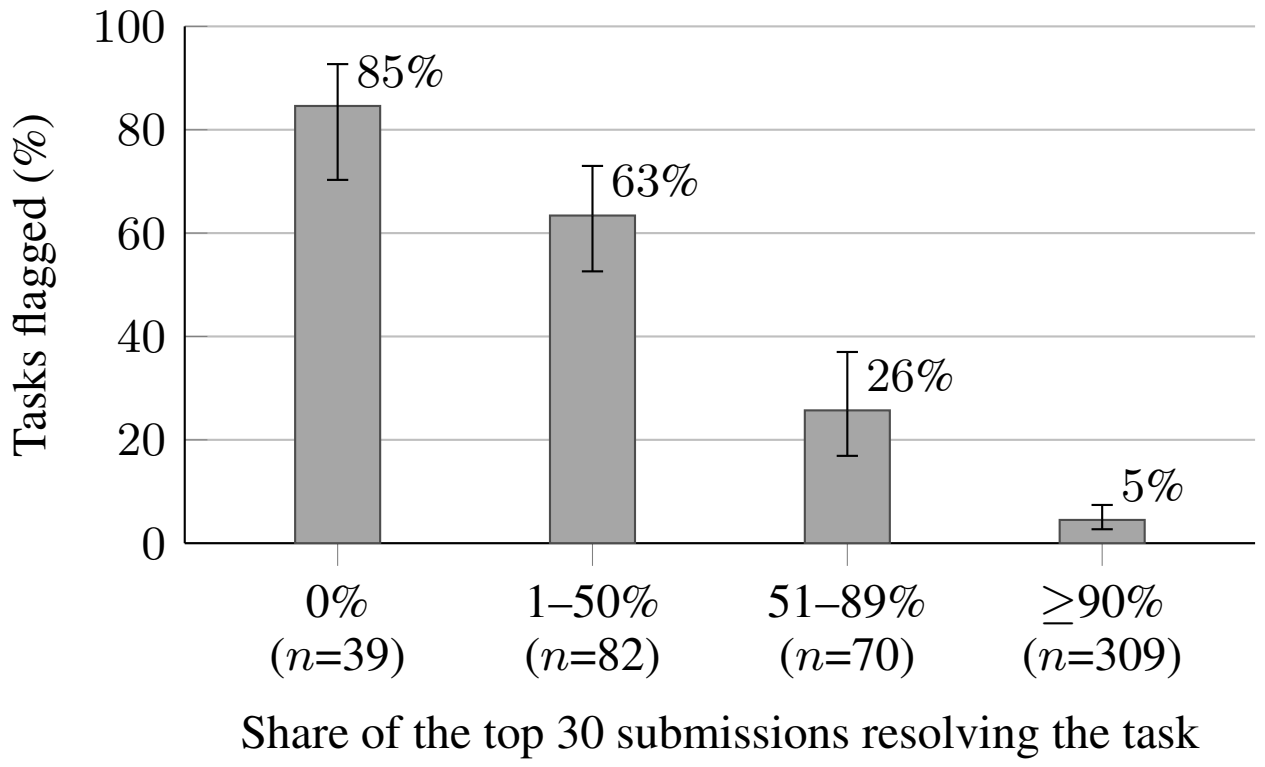
\begin{figure}[H]
\centering
\begin{minipage}{0.62\linewidth}\begin{tikzpicture}
\begin{axis}[ybar, bar width=16pt, width=\linewidth, height=5cm,
  label style={font=\small}, tick label style={font=\footnotesize},
  ymin=0, ymax=100, enlarge x limits=0.2, ylabel={Tasks flagged (\%)},
  xlabel={Share of the top 30 submissions resolving the task},
  xtick={1,2,3,4}, xticklabels={{0\%\\($n$=39)},{1--50\%\\($n$=82)},{51--89\%\\($n$=70)},{$\geq$90\%\\($n$=309)}}, xticklabel style={align=center},
  ymajorgrids, axis lines*=left, nodes near coords, nodes near coords style={font=\footnotesize, anchor=south west, text=black},
  point meta=y, nodes near coords={\pgfmathprintnumber[fixed,precision=0]{\pgfplotspointmeta}\%}]
\addplot+[fill=black!35, draw=black!70, error bars/.cd, y dir=both, y explicit,
  error bar style={black}] coordinates {
  (1,84.6) += (0,8.1) -= (0,14.3)
  (2,63.4) += (0,9.6) -= (0,10.8)
  (3,25.7) += (0,11.3) -= (0,8.8)
  (4,4.5) += (0,2.9) -= (0,1.8)
};
\end{axis}
\end{tikzpicture}\end{minipage}
\caption{Share of Verified tasks flagged, by how many of the top \topN{} public submissions resolve the task. Bars show Wilson 95\% intervals.}
\label{fig:resolve}
\end{figure}

\section{Flag rate by difficulty}
\label{app:difficulty}

\begin{table}[H]
\caption{Verified tasks by OpenAI's difficulty rating (annotators' estimate of time to fix). The three tasks rated over 4 hours are included in ``$>$1 hour''. Mean resolve rate is over the top \topN{} submissions.}
\label{tab:difficulty}
\centering\small
\fit{\begin{tabular}{lrlrr}
\toprule
Difficulty & Tasks & Flagged & \multicolumn{2}{c}{Mean resolve rate} \\
 & & & flagged & not flagged \\
\midrule
$<$15 min fix & 194 & 13\% (26/194; 9--19\%) & 41\% & 93\% \\
15 min - 1 hour & 261 & 25\% (65/261; 20--30\%) & 32\% & 85\% \\
$>$1 hour & 45 & 58\% (26/45; 43--71\%) & 16\% & 68\% \\
\bottomrule
\end{tabular}
}
\end{table}

\section{Flip rate by resolve band}
\label{app:bands}

\begin{table}[H]
\caption{Share of regraded patches that flip, by how many of the top \topN{} submissions resolve the task under the original tests.}
\label{tab:execbands}
\centering\small
\fit{\begin{tabular}{lll}
\toprule
Resolved by & Flagged & Control, forced \\
\midrule
0\% of top 30 & 73\% (545/745) & 0\% (0/30) \\
1-50\% of top 30 & 72\% (696/969) & 42\% (75/178) \\
51-89\% of top 30 & 69\% (82/118) & 26\% (32/125) \\
$\geq$90\% of top 30 & 62\% (8/13) & 30\% (13/44) \\
\bottomrule
\end{tabular}
}
\end{table}

\section{Examples of flipped patches}
\label{app:flips}

Table~\ref{tab:flips} gives three flagged tasks. In each, the original test fixes a detail the issue leaves open or contradicts, and the flipped patch makes a different choice that the issue supports.

\begin{table}[H]
\caption{Flipped patches on three flagged tasks (flips out of regraded candidates in parentheses).}
\label{tab:flips}
\centering\small
\begin{tabular}{@{}p{0.15\linewidth}>{\raggedright\arraybackslash}p{0.22\linewidth}>{\raggedright\arraybackslash}p{0.30\linewidth}>{\raggedright\arraybackslash}p{0.24\linewidth}@{}}
\toprule
Task & The issue asks for & Original test and relaxed test & A patch that flips \\
\midrule
\code{django-11790} (\exMaxlength) & The username field of \code{AuthenticationForm} should render the \code{maxlength} HTML attribute again. & Original: \code{widget.\allowbreak attrs['maxlength']} equals the integer 255. Relaxed: the rendered field contains \code{maxlength="255"}. & Stores the string \code{'255'}, as Django's own \code{CharField} does. The rendered HTML is the same as with the reference fix. \\
\addlinespace
\code{django-12308} (\exJson) & A \code{JSONField} shown read-only in the admin should display valid JSON, not a Python dict. The issue proposes calling the field's \code{prepare\_value}. & Original: besides the valid-JSON cases, a dict with a tuple key, which cannot be stored as JSON, must display as its Python \code{repr}. Relaxed: that case is removed. & Calls the form field's \code{prepare\_value}, as the issue proposes. It raises \code{TypeError} on the tuple-key dict. \\
\addlinespace
\code{pylint-4661} (\exPylint) & Stop storing data in \code{\textasciitilde/.pylint.d} and follow the XDG specification, which puts data in \code{\textasciitilde/.local/\allowbreak share/\allowbreak <program>}. & Original: \code{PYLINT\_HOME} equals \code{appdirs.\allowbreak user\_cache\_dir(\allowbreak "pylint")}, the cache directory. Relaxed: an absolute path that names pylint and is neither \code{\textasciitilde/.pylint.d} nor directly in the home directory. & Uses \code{\textasciitilde/.local/\allowbreak share/pylint}, the location the issue names. No top-\topN{} submission resolves the original task. \\
\bottomrule
\end{tabular}
\end{table}

\section{Scores under relaxed tests}
\label{app:agents}

\begin{table}[H]
\caption{The top \topN{} Verified submissions if every flip on a flagged task counted as resolved (Section~\ref{sec:exec}).}
\centering\scriptsize
\fit{\begin{tabular}{lrrrrr}
\toprule
Submission & Resolved & Flips & After & Rank & Rank after \\
\midrule
\texttt{20251205\_sonar-foundation-agent\_claude-opus-4-5} & 396 & 41 & 437 & 1 & 3 \\
\texttt{20251215\_livesweagent\_claude-opus-4-5} & 396 & 38 & 434 & 1 & 4 \\
\texttt{20250928\_trae\_doubao\_seed\_code} & 394 & 45 & 439 & 3 & 1 \\
\texttt{20251127\_openhands\_claude-opus-4-5} & 388 & 44 & 432 & 4 & 5 \\
\texttt{20251120\_livesweagent\_gemini-3-pro-preview} & 387 & 43 & 430 & 5 & 7 \\
\texttt{20250804\_epam-ai-run-claude-4-sonnet} & 384 & 38 & 422 & 6 & 10 \\
\texttt{20250902\_atlassian-rovo-dev} & 384 & 54 & 438 & 6 & 2 \\
\texttt{20250819\_ACoder} & 382 & 37 & 419 & 8 & 14 \\
\texttt{20250901\_warp} & 378 & 54 & 432 & 9 & 5 \\
\texttt{20260217\_mini-v2.0.0\_claude-4-6-opus} & 378 & 45 & 423 & 9 & 8 \\
\texttt{20250612\_trae} & 376 & 39 & 415 & 11 & 16 \\
\texttt{20250731\_harness\_ai} & 374 & 40 & 414 & 12 & 18 \\
\texttt{20251103\_sonar-foundation-agent\_claude-sonnet-4-5} & 374 & 49 & 423 & 12 & 8 \\
\texttt{20250720\_Lingxi-v1.5\_claude-4-sonnet-20250514} & 373 & 37 & 410 & 14 & 21 \\
\texttt{20250915\_JoyCode} & 373 & 33 & 406 & 14 & 24 \\
\texttt{20250603\_Refact\_Agent\_claude-4-sonnet} & 372 & 41 & 413 & 16 & 19 \\
\texttt{20251124\_mini-v1.16.0\_claude-opus-4-5-20251101} & 372 & 48 & 420 & 16 & 12 \\
\texttt{20251118\_mini-v1.15.0\_gemini-3-pro-preview-20251118} & 371 & 49 & 420 & 18 & 12 \\
\texttt{20251103\_SalesforceAIResearch\_SAGE\_OpenHands} & 369 & 53 & 422 & 19 & 10 \\
\texttt{20250522\_tools\_claude-4-opus} & 366 & 49 & 415 & 20 & 16 \\
\texttt{20251021\_SalesforceAIResearch\_SAGE\_bash\_only} & 365 & 48 & 413 & 21 & 19 \\
\texttt{20250522\_tools\_claude-4-sonnet} & 362 & 45 & 407 & 22 & 23 \\
\texttt{20250807\_openhands\_gpt5} & 359 & 50 & 409 & 23 & 22 \\
\texttt{20251211\_mini-v1.17.2\_gpt-5.2-2025-12-11-high} & 359 & 60 & 419 & 23 & 14 \\
\texttt{20250710\_bloop} & 356 & 41 & 397 & 25 & 28 \\
\texttt{20250715\_qodo\_command} & 356 & 46 & 402 & 25 & 25 \\
\texttt{20251014\_Lingxi\_kimi\_k2} & 356 & 34 & 390 & 25 & 29 \\
\texttt{20250623\_warp} & 355 & 46 & 401 & 28 & 27 \\
\texttt{20250611\_moatless\_claude-4-sonnet-20250514} & 354 & 48 & 402 & 29 & 25 \\
\texttt{20250519\_trae} & 353 & 36 & 389 & 30 & 30 \\
\bottomrule
\end{tabular}
}
\end{table}

\section{Flag rates by repository}
\label{app:repos}

\begin{table}[H]
\caption{SWE-bench Verified by repository.}
\centering\small
\fit{\begin{tabular}{lrlrr}
\toprule
Repository & Tasks & Flagged & Score 3 & Score 4 \\
\midrule
django/django & 231 & 20\% (46/231; 15--26\%) & 33 & 13 \\
sympy/sympy & 75 & 31\% (23/75; 21--42\%) & 17 & 6 \\
sphinx-doc/sphinx & 44 & 25\% (11/44; 15--39\%) & 9 & 2 \\
matplotlib/matplotlib & 34 & 26\% (9/34; 15--43\%) & 3 & 6 \\
scikit-learn/scikit-learn & 32 & 25\% (8/32; 13--42\%) & 4 & 4 \\
astropy/astropy & 22 & 27\% (6/22; 13--48\%) & 5 & 1 \\
pydata/xarray & 22 & 23\% (5/22; 10--43\%) & 3 & 2 \\
pytest-dev/pytest & 19 & 16\% (3/19; 6--38\%) & 1 & 2 \\
pylint-dev/pylint & 10 & 50\% (5/10; 24--76\%) & 2 & 3 \\
psf/requests & 8 & 0\% (0/8; 0--32\%) & 0 & 0 \\
mwaskom/seaborn & 2 & 50\% (1/2; 9--91\%) & 1 & 0 \\
pallets/flask & 1 & 0\% (0/1; 0--79\%) & 0 & 0 \\
\bottomrule
\end{tabular}
}
\end{table}

\begin{table}[H]
\caption{SWE-bench Pro (public set) by repository.}
\centering\small
\fit{\begin{tabular}{lrlrr}
\toprule
Repository & Tasks & Flagged & Score 3 & Score 4 \\
\midrule
ansible/ansible & 96 & 43\% (41/96; 33--53\%) & 27 & 14 \\
internetarchive/openlibrary & 91 & 21\% (19/91; 14--30\%) & 10 & 9 \\
flipt-io/flipt & 85 & 42\% (36/85; 32--53\%) & 12 & 24 \\
qutebrowser/qutebrowser & 79 & 30\% (24/79; 21--41\%) & 17 & 7 \\
gravitational/teleport & 76 & 46\% (35/76; 35--57\%) & 9 & 26 \\
protonmail/webclients & 65 & 34\% (22/65; 24--46\%) & 16 & 6 \\
future-architect/vuls & 62 & 35\% (22/62; 25--48\%) & 11 & 11 \\
navidrome/navidrome & 57 & 35\% (20/57; 24--48\%) & 10 & 10 \\
element-hq/element-web & 56 & 45\% (25/56; 32--58\%) & 19 & 6 \\
NodeBB/NodeBB & 44 & 23\% (10/44; 13--37\%) & 10 & 0 \\
tutao/tutanota & 20 & 20\% (4/20; 8--42\%) & 2 & 2 \\
\bottomrule
\end{tabular}
}
\end{table}

\section{API cost}
\label{app:cost}

\begin{table}[H]
\caption{Tokens and cost of the model calls behind each result, at standard list prices.}
\label{tab:cost}
\centering\small
\fit{\begin{tabular}{lrrrr}
\toprule
Run & Calls & Input tokens (M) & Output tokens (M) & Cost \\
\midrule
Judge panel, all tasks & 2,686 & 13.37 & 3.32 & \$94 \\
Ablation: identical rerun & 276 & 1.23 & 0.31 & \$9 \\
Ablation: with reference fix & 276 & 1.93 & 0.31 & \$11 \\
Ablation: baseline prompt & 276 & 0.79 & 0.19 & \$6 \\
Relaxer, with findings & 110 & 0.56 & 0.23 & \$7 \\
Relaxer, no findings (both groups) & 199 & 0.86 & 0.25 & \$8 \\
Relaxer, forced & 94 & 0.38 & 0.11 & \$4 \\
\midrule
Total & 3,917 & 19.12 & 4.71 & \$138 \\
\bottomrule
\end{tabular}
}
\end{table}

\end{document}